\documentclass[aps,prb,twocolumn,longbibliography]{revtex4-1}

\usepackage{latexsym}
\usepackage{graphicx} 
\usepackage{epstopdf}
\usepackage{amsmath}  
\usepackage{amssymb}
\usepackage{comment}
\usepackage{cases}

\DeclareMathOperator*{\SumInt}{%
\mathchoice%
  {\ooalign{$\displaystyle\sum$\cr\hidewidth$\displaystyle\int$\hidewidth\cr}}
  {\ooalign{\raisebox{.14\height}{\scalebox{.7}{$\textstyle\sum$}}\cr\hidewidth$\textstyle\int$\hidewidth\cr}}
  {\ooalign{\raisebox{.2\height}{\scalebox{.6}{$\scriptstyle\sum$}}\cr$\scriptstyle\int$\cr}}
  {\ooalign{\raisebox{.2\height}{\scalebox{.6}{$\scriptstyle\sum$}}\cr$\scriptstyle\int$\cr}}
}

\begin{document}

\title{Stable high-temperature paramagnons in a three-dimensional antiferromagnet near  quantum criticality: Application to $\boldsymbol{\mathrm{TlCuCl_3}}$}

\author{Maciej Fidrysiak}
 \email{maciej.fidrysiak@uj.edu.pl}
\author{J{\'o}zef Spa{\l}ek}%
 \email{jozef.spalek@uj.edu.pl}
\affiliation{Marian Smoluchowski Institute of Physics, Jagiellonian University, ul. {\L}ojasiewicza 11, 30-348 Krak{\'o}w, Poland }%


\date{\today}

\begin{abstract}
The complete set of hallmarks of the three-dimensional antiferromagnet near the quantum critical point has been recently observed in the spin dimer compound $\mathrm{TlCuCl_3}$. Nonetheless, the mechanism, responsible for several distinct features of the  experimental data, has remained a puzzle, namely: (i) the paramagnons exhibit remarkable robustness to thermal damping and are stable up to high temperatures, where $k_B T$ is comparable with the excitation energy, (ii) the width to mass ratios of the high-temperature paramagnons are, within the error bars, equal to that of the low-temperature amplitude (or Higgs) mode. We propose such a mechanism and identify two principal factors, contributing to the scaling between width to mass ratios of the paramagnon and the amplitude mode: (i) the emergence of the thermal mass scale reorganizing the paramagnon decay processes, and (ii) substantial renormalization of the multi-magnon interactions by thermal fluctuations. The study is carried out for the general case of a $D=3 + 1$ quantum antiferromagnet within the framework of the $\varphi^4$ model using the hybrid Callan-Symanzik + Wilson thermal renormalization group method. Our approach is tested by demonstrating a good quantitative agreement with available experimental data across the phase diagram of $\mathrm{TlCuCl_3}$.
\end{abstract}


\maketitle

\section{Introduction}
\label{sec:introduction}

\begin{figure}%
\centering
\parbox{0.36\textwidth}{\centering\includegraphics[width = 0.36\textwidth]{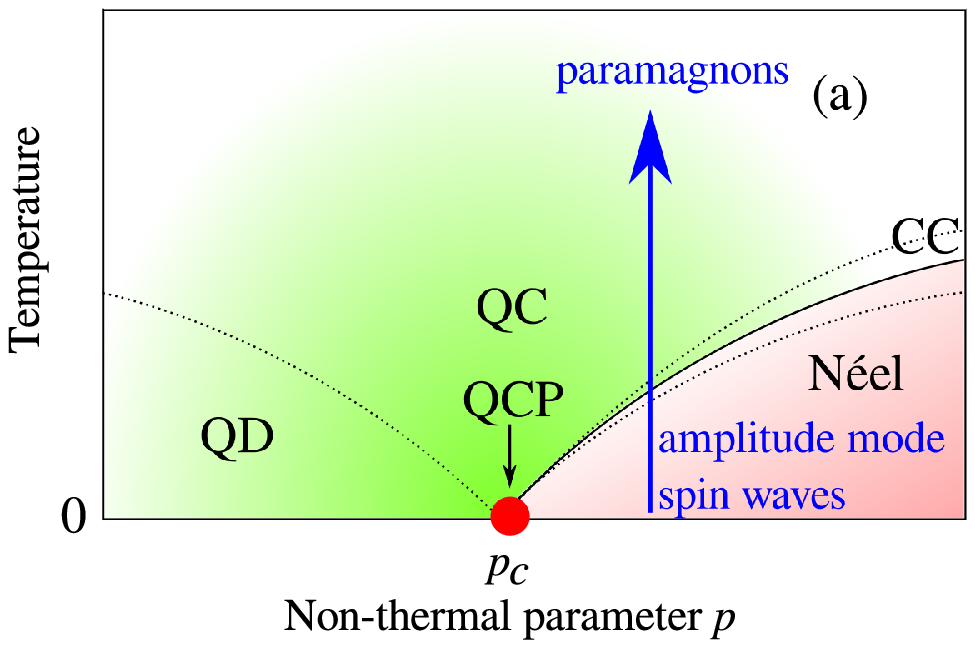}}%
\qquad
\parbox{0.40\textwidth}{\includegraphics[width = 0.40\textwidth]{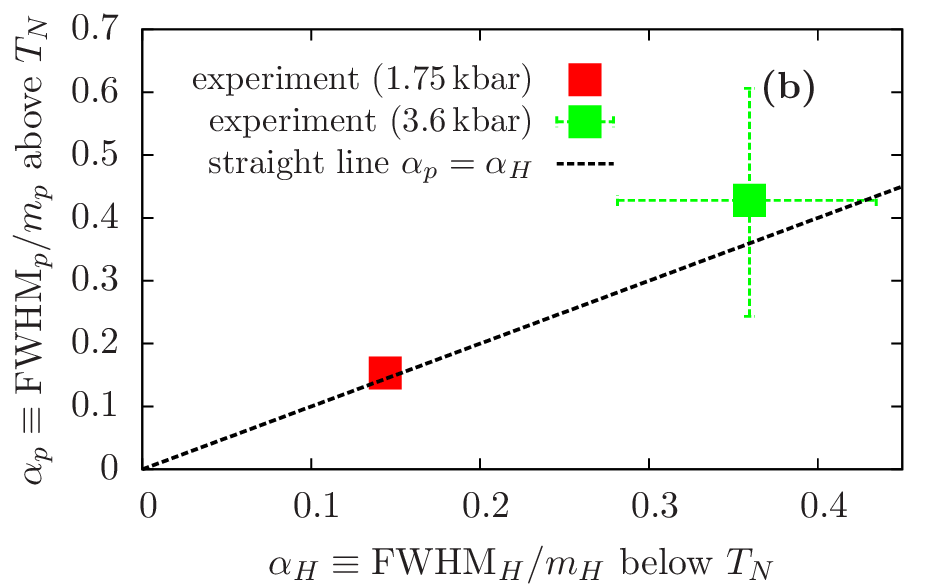}}%
\caption{(Color online) (a) A schematic phase diagram of a three-dimensional antiferromagnet near the quantum critical point: N\'{e}el -- antiferromagnetic phase, QD -- quantum disordered, QC -- quantum critical, CC -- classical critical. The dotted lines mark crossovers. As one moves along the vertical arrow, the magnetic excitations evolve from spin waves and the amplitude mode at low-temperature to the high-temperature paramagnons. (b) The scaling between the width to mass ratio for the high-temperature paramagnons and that for the low-temperature amplitude mode. Squares are experimental data of Ref.~\citenum{NatPhys10Merchant} and the dashed straight line $\alpha_p = \alpha_H$ is guide to the eye. The high-temperature data have been collected at $\approx 11.5$ and $13\,\mathrm{K}$ for $p = 1.75$ and $3.6\,\mathrm{kbar}$, respectively. The low-temperature amplitude mode width to mass ratio has been measured at $\approx 1.8\,\mathrm{K}$ in both cases.} \label{fig:introduction}
\end{figure}

Phase diagram of a quantum antiferromagnet in three spatial dimensions involves a quantum critical point (QCP) which separates the N{\'e}el phase from the quantum disordered state (QD) at zero temperature [cf. Fig.~1(a)]. Its central part is occupied by the quantum critical (QC) regime indicating non-trivial interplay between quantum and thermal fluctuations.\cite{PhysRevB39Chakravarty} The transitions between these phases are governed by temperature, as well as by non-thermal parameters that couple to zero-point fluctuations. As one moves along the zero-temperature line, the quasiparticles evolve from the massive excitations in the QD phase to massless spin-wave modes in the N\'{e}el state. Additionally, the amplitude (or Higgs) mode, associated with spin fluctuations directed along the ordered moments, is expected on the ordered side of the phase diagram.\cite{PhysRevB46Affleck} Recently achieved control of dimerized antiferromagnets $\mathrm{TlCuCl_3}$ and $\mathrm{KCuCl_3}$ near the pressure-induced quantum phase transition allows for probing all phases depicted in Fig.~\ref{fig:introduction}(a).\cite{PhysRevLett100Ruegg,JPhysConfSeries400Kuroe,NatPhys10Merchant}

Close to the critical pressure $p_c \approx 1.07\,\mathrm{kbar}$, stable amplitude modes, characterized by full width at half maximum to mass ratios as low as $\alpha_H \equiv \mathrm{FWHM}_{H}/m_{H} \approx 0.2$, have been observed in $\mathrm{TlCuCl_3}$.\cite{PhysRevLett100Ruegg,NatPhys10Merchant} The width to mass ratio of the paramagnons $\alpha_p \equiv \mathrm{FWHM}_p / m_p$ above the N\'{e}el temperature $T_N$ turns out to be small in this compound as well. Namely, within the error bars, the equality $\alpha_p(T \gg T_N) \approx \alpha_H(T \ll T_N)$ holds for various values of pressure. In Fig.~\ref{fig:introduction}(b) this scaling for $\mathrm{TlCuCl_3}$ is depicted along with the straight line $\alpha_p = \alpha_H$ as guide to the eye. Since the paramagnons have been probed at high temperatures (i.e., for $k_B T$ comparable with the excitation energies), the latter relation implies a remarkably small influence of thermal disorder on broadening of magnetic excitations.\cite{NatPhys10Sushkov} In this paper we propose a mechanism protecting the paramagnons from thermal damping and identify two principal factors responsible for the behavior depicted in Fig.~\ref{fig:introduction}(b): (i) the emergence of the thermal mass scale $m_T \propto \sqrt{\lambda} \cdot T$, where $\lambda$ denotes the properly normalized coupling constant (see the discussion below). This particular $\lambda$-dependence reorganizes perturbation theory in such a way that $\alpha_p(T \gg T_N)$  becomes formally of the same order as $\alpha_H(T \ll T_N)$ and thus allows for the linear scaling between these quantities, (ii) a substantial downward renormalization of $\lambda$ by thermal fluctuations, which, in turn, leads to reduction of $\alpha_p(T \gg T_N)$. We demonstrate that the latter effect needs to be included to obtain the scaling with the correct proportionality factor. 

The study is performed for the general case of three-dimensional ($D = 3 + 1$) antiferromagnet near the quantum critical point, within the framework of the effective $\varphi^4$ theory.  We employ the hybrid zero-temperature and Wilson thermal renormalization group method that allows to interpolate between the quantum- and classical-critical behaviors. It is thus suitable for a detailed comparison with experiments on $\mathrm{TlCuCl_3}$.

The paper is organized as follows. In Sec.~\ref{section:widths} we derive the widths of magnetic excitations below and above the N\'{e}el temperature. In Sec.~\ref{section:RG} we describe the renormalization group procedure. In Sec.~\ref{section:main_result} we propose the mechanism responsible for the scaling between $\alpha_p$ and $\alpha_H$. Finally, in Sec.~\ref{section:experiment}, we perform a comparison with the neutron scattering data on $\mathrm{TlCuCl_3}$ across the quantum critical phase diagram. The study is summarized and related to other recent approaches in Sec.~\ref{sec:summary}. The Appendices~\ref{appendix:widths_of_excitations}-\ref{appendix:RG_equations}
provide methodological details of the analysis.

\section{$\boldsymbol{\varphi^4}$ model and width of the magnetic excitations}
\label{section:widths}

We are concerned with the long-wavelength characteristics of magnetic excitations close to the magnetic quantum critical point. In this regime the low-energy physics is expected to be insensitive to the microscopic details of the compound under consideration and can be reliably studied within an effective model approach. Here we employ the $\varphi^4$ theory rather than a microscopic spin-dimer model, previously applied to $\mathrm{TlCuCl_3}$ and $\mathrm{KCuCl_3}$.\cite{PhysRevB69Matsumoto} The magnetic excitations reflect then different fluctuation modes of the three-component ($N = 3$) local N\'{e}el order parameter $\boldsymbol{\varphi}$ whose dynamics is governed by the Lagrangian

\begin{align}
\mathcal{L} = & \frac{1}{2} (\nabla \boldsymbol{\varphi})^2 + \frac{1}{2} m^2 \boldsymbol{\varphi}^2 + \frac{1}{4} \lambda \left(\boldsymbol{\varphi}^2\right)^2 \nonumber \\ &+ \frac{1}{2} \delta m^2 (\boldsymbol{\varphi})^2 + \frac{1}{4} \delta \lambda \left(\boldsymbol{\varphi}^2\right)^2. \label{eq:lagrangian}
\end{align}

\noindent
Here $m$ and $\lambda$ denote the mass parameter and the coupling constant, respectively, whereas $\delta m$ and $\delta \lambda$ are the counter-terms introduced to cancel off the short-wavelength divergences. We work in the natural units by setting to unity the spin-wave velocity ($c = 1$), as well as the Planck's ($\hbar = 1$) and the Boltzmann ($k_B = 1$) constants. Both imaginary time $\tau$ and spatial coordinates $x, y, z$ have then the dimension $\mathrm{energy}^{-1}$, mass $m$, field $\boldsymbol{\varphi}$, and temperature are measured in the units of energy, while $\lambda$ is dimensionless. The four-gradient symbol takes then the form  $\nabla = (\partial_\tau, \partial_x, \partial_y, \partial_z)$. 

\begin{figure}%
\centering
\parbox{0.49\textwidth}{\includegraphics[width = 0.49\textwidth]{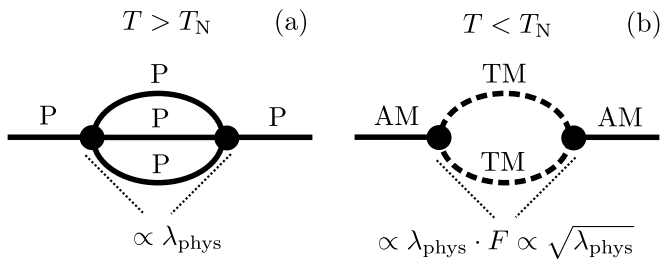}}%
\caption{Leading-order processes yielding broadening of (a) paramagnons above the N\'{e}el temperature, (b) amplitude mode below the N\'{e}el temperature. Labeling of the lines: P -- paramagnon, AM -- amplitude mode, TM -- transverse mode. } \label{fig:decay_processes}
\end{figure}

In the disordered phase, where the physical (``dressed'' with quantum and thermal corrections) mass parameter squared $m_\mathrm{phys}^2$ is positive, all the three fluctuation modes of the local order parameter $\boldsymbol{\varphi}$ are equivalent paramagnons of mass $m_p = m_\mathrm{phys}$. In this case, the leading order contribution to broadening of the magnetic excitations (in the sense of perturbative expansion in the effective coupling constant $\lambda_\mathrm{phys}$) comes out directly from the quartic term $\frac{1}{4} \lambda_\mathrm{phys} \cdot (\boldsymbol{\varphi}^2)^2 $ and is given by the sunset diagram, shown in Fig.~\ref{fig:decay_processes}(a). To derive explicit form of $m^2_\mathrm{phys}$ and $\lambda_\mathrm{phys}$ one needs to specify computational scheme which is detailed below.

In the N\'{e}el phase ($m^2_\mathrm{phys} < 0$), the order parameter acquires a non-zero expectation value, breaking the spin-rotational symmetry of the system. Without loss of generality one can take $\left<{\boldsymbol{\varphi}}\right> = (0, \ldots, 0, F)$, where $F^2 \approx {-m^2_\mathrm{phys}/\lambda_\mathrm{phys}}$. As a consequence of symmetry breaking, the longitudinal and transverse fluctuations  (defined, respectively, as $\sigma \equiv \varphi^N - F$ and $\pi^i \equiv \varphi^i$ for $i = 1, \ldots, N - 1$, with $N = 3$) are no longer equivalent. Also, qualitatively new interactions between $\sigma$ and $\boldsymbol{\pi}$ emerge, including the three-point vertex $V_{\sigma\boldsymbol{\pi}\boldsymbol{\pi}} = \lambda_\mathrm{phys} F \cdot \sigma \boldsymbol{\pi}^2$. The leading order process contributing to the amplitude mode damping is now generated by $V_{\sigma\boldsymbol{\pi}\boldsymbol{\pi}}$ and is represented by the one-loop diagram shown in Fig.~\ref{fig:decay_processes}(b). 

By evaluating the diagrams of Fig.~\ref{fig:decay_processes} (cf. Appendix~\ref{appendix:widths_of_excitations} and Ref.~\citenum{PhysRevD53Wang}.), we find the full width to mass ratio of the paramagnon above $T_N$

\begin{align}
\alpha_p \equiv &\frac{\mathrm{FWHM}_p}{m_p} = \frac{3\lambda^2_\mathrm{phys} (N + 2)}{32 \pi^3} \cdot \frac{T^2}{m_p^2} \cdot \mathrm{Li}_2(\mathrm{e}^{-m_p/T}) \label{eq:paramagnon_stability}
\end{align}

\noindent
and that of the amplitude mode below $T_N$

\begin{align}
\alpha_H \equiv &\frac{\mathrm{FWHM}_H}{m_H} = \frac{\lambda_\mathrm{phys} (N - 1)}{16 \pi} \cdot [1 + 2\cdot  n(m_H/2)], \label{eq:higgs_stability}
\end{align}

\noindent
respectively. Here $\mathrm{Li}_2(x) = \int_{1}^{x} dt \frac{\ln(t)}{1 - t}$ denotes the dilogarithm.

\section{Determinantion of the effective parameters}
\label{section:RG}

The stability parameters of the paramagnon and the amplitude mode, $\alpha_p$ and $\alpha_H$ [Eqs.~(\ref{eq:paramagnon_stability})-(\ref{eq:higgs_stability})], are formally of different order in the effective coupling constant $\lambda_\mathrm{phys}$, i.e., $\alpha_p = O(\lambda^2_\mathrm{phys})$ and $\alpha_H = O(\lambda_\mathrm{phys})$. The proportionality between $\alpha_p$ and $\alpha_H$, established experimentally, is hence inconsistent with the casual perturbation theory. Here we perform a non-perturbative resummation of both quantum and thermal fluctuations by the hybrid renormalization group (RG) method, which allows to track down higher-order effects contributing to the ratio $\alpha_p / \alpha_H$. Additionally, the latter approach is applicable close both to the quantum and classical transitions and hence provides a unified picture of the quantum critical phase diagram. This property makes it suitable for a global comparison with experiments on dimerized antiferromagnets, where the magnetic excitations have probed in all phases depicted in Fig.~\ref{fig:introduction}(a). Here we sketch the derivation of the renormalization group equations and provide the technical details in Appendix~\ref{appendix:RG_equations}.

The hybrid RG relies on the two scales: (i) renormalization scale $\mu$, introduced by the procedure of subtracting the short-range divergences, and (ii) infrared momentum cutoff to the thermal fluctuations $\Lambda$, implemented by redefining the Bose occupation factors $n(E_\mathbf{k}) \equiv (\exp(E_\mathbf{k}/T) - 1)^{-1}$ as $n(E_\mathbf{k}) \leftrightarrow n_\Lambda(E_\mathbf{k}) \equiv n(E_\mathbf{k}) \cdot \theta(|\mathbf{k}| - \Lambda)$, where $\theta$ is the Heaviside step function and $E_\mathbf{k}$ is the energy of the excitation of wavevector $\mathbf{k}$. In the $\Lambda \rightarrow 0$ limit the full factor $n(E)$ is recovered, whereas for $\Lambda \rightarrow \infty$ all thermally excited modes are suppressed ($n_{\Lambda\rightarrow \infty} = 0$). 

The limit $\Lambda \rightarrow \infty$ corresponds hence to $T \rightarrow 0$, where the scale dependence of the running mass $m_\mu$ and the coupling constant $\lambda_\mu$ can be found by solving the $T = 0$ Callan-Symanzik equations

\begin{numcases}{}
\mu \frac{\partial \lambda_\mu}{\partial \mu} = \frac{2 (N + 8)}{(4 \pi)^2} \lambda^2_\mu, \label{eq:RG_T0_lambda} \\
\frac{\mu}{m^2} \frac{\partial m^2_\mu}{\partial \mu} = \frac{2 (N + 2)}{(4 \pi)^2} \lambda_\mu. \label{eq:RG_T0_mass}
\end{numcases}

\noindent
The physically relevant scale is provided by the paramagnon mass in the disordered phase $m_p = m_\mathrm{phys}$ and by the amplitude mode mass in the N\'{e}el state $m_H =  (2 \cdot |m_\mathrm{phys}^2|)^{1/2}$. Since Eqs.~(\ref{eq:RG_T0_lambda})-(\ref{eq:RG_T0_mass}) describe essentially mean field behavior with weakly scale-dependent logarithmic corrections, we are allowed to take $\mu = |m_\mathrm{phys}^2|^{1/2}$ in both phases without imposing significant error so that $m_{\mathrm{phys}}^2(T = 0) = m^2_\mu \mid_{\mu = \sqrt{|m_\mathrm{phys}^2|}}$ and $\lambda_{\mathrm{phys}}(T = 0) = \lambda_\mu \mid_{\mu = \sqrt{|m_\mathrm{phys}^2|}}$. 

Once the zero-temperature parameters are known, the finite-temperature effects can be incorporated by progressively integrating out thermal fluctuations and moving from $\Lambda \rightarrow \infty$ to $\Lambda = 0$. The initial conditions at $\Lambda = \infty$ are now given by the physical zero-temperature quantities. This step is equivalent to the Wilson thermal renormalization group method, previously discussed within the real-time\cite{NuclPhysB472Attanasio} and imaginary-time\cite{PhysRevD52Liao} formulation of thermal theory, and leads to the following set of differential equations equations with respect to  $\Lambda$:

\begin{numcases}{}
\Lambda \frac{{d}\lambda_\Lambda}{{d}\Lambda}  = (N - 1) \lambda_\Lambda^2 \mathcal{I}_\Lambda^{''}(m_{\perp, \Lambda}^2) + 9 \lambda_\Lambda^2 \mathcal{I}_\Lambda^{''}(m_{||, \Lambda}^2), \label{eq:TRG_lambda}  \\
\Lambda \frac{{d}  m^2_\Lambda}{{d}\Lambda} = (N - 1) \lambda_\Lambda \mathcal{I}_\Lambda^{'}(m_{\perp, \Lambda}^2) + 3 \lambda_\Lambda \mathcal{I}_\Lambda^{'}(m_{||}^2)  \nonumber \\ \hspace{4 em} - (N - 1) \lambda^2 F_\Lambda^2 \mathcal{I}_\Lambda^{''}(m_{\perp, \Lambda}^2) \nonumber \\ \hspace{4em}- 9 \lambda^2 F_\Lambda^2 \mathcal{I}_\Lambda^{''}(m_{||, \Lambda}^2),\label{eq:TRG_mass}
\end{numcases}

\noindent
where $F^2_\Lambda = \max(0, -m^2_\Lambda/\lambda_\Lambda)$ is the square of the antiferromagnetic order parameter, $m^2_{\perp, \Lambda} \equiv m^2_\Lambda + \lambda_\Lambda F_\Lambda^2$, and $m_{||, \Lambda}^2 \equiv m^2_\Lambda + 3 \lambda_\Lambda F^2_\Lambda$. The temperature enters through the expressions

\begin{align}
\mathcal{I}_\Lambda^{'}(M^2) = &-\frac{\Lambda^3}{2\pi^2} \frac{n(\sqrt{M^2 + \Lambda^2})}{\sqrt{M^2 + \Lambda^2}}, \label{eq:I'}\\
\mathcal{I}_\Lambda^{''}(M^2) = &-\frac{\Lambda^3}{2\pi^2} \frac{d}{d M^2} \left[ \frac{n(\sqrt{M^2 + \Lambda^2})}{\sqrt{M^2 + \Lambda^2}} \right]. \label{eq:I''}
\end{align}

In the next section we analyze the solutions of the above equations to determine the stability of the high-temperature paramagnons.

\section{Stable high-temperature paramagnons}
\label{section:main_result}

\begin{figure}%
\centering
\parbox{0.39\textwidth}{\includegraphics[width = 0.39\textwidth]
{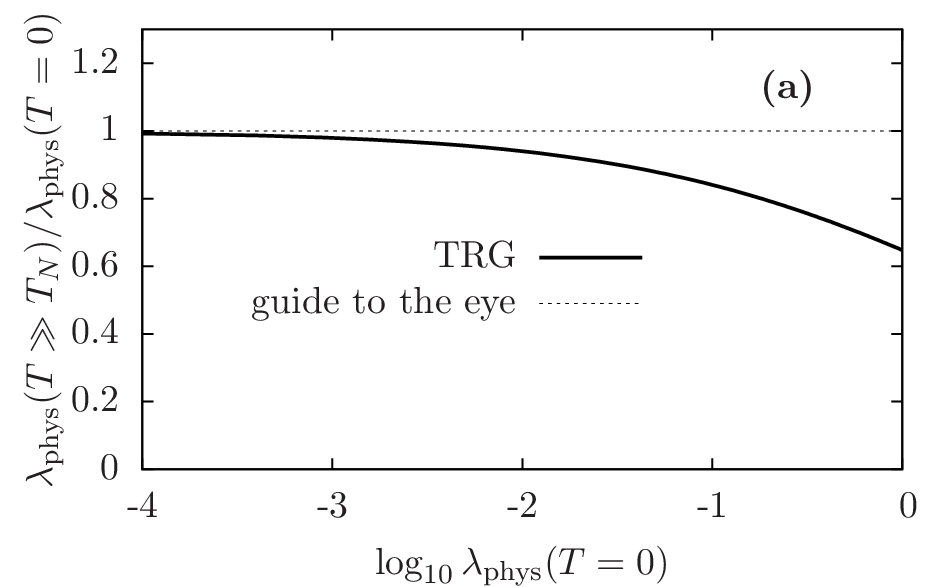}}%

\parbox{0.39\textwidth}{\includegraphics[width = 0.39\textwidth]{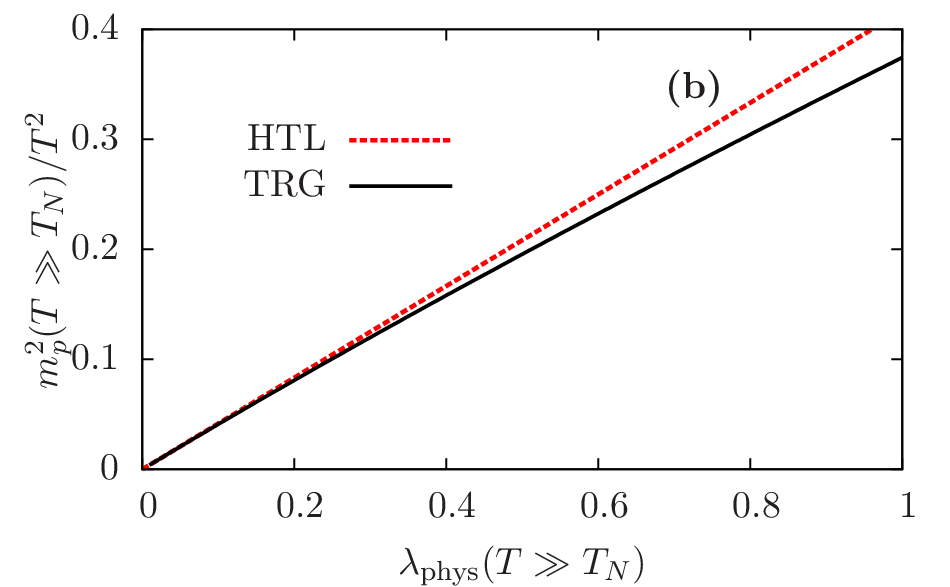}}%
\caption{(Color online) (a) Ratio of the high- to the low-temperature coupling constants as a function of the low-temperature coupling $\lambda_\mathrm{phys}(T = 0)$. (b) Normalized square of the high-temperature paramagnon mass as a function of the high-temperature coupling constant. TRG -- thermal renormalization group. HTL -- result of the resummation of the hard thermal loops (see the text).} \label{fig:thermal_mass}
\end{figure}

We now turn to the main result of the paper and explain the relation between $\alpha_p(T \gg T_N)$ and $\alpha_H(T \ll T_N)$. For that purpose, we consider the coupling constant $\lambda_\mathrm{phys}(T = 0)$  as a free variable and calculate both $\alpha_H(T \ll T_N)$ and $\alpha_p(T \gg T_N)$ as a function of $\lambda_\mathrm{phys}(T = 0)$. 

The low-temperature amplitude mode stability parameter $\alpha_H(T \ll T_N)$ is uniquely determined by $\lambda_\mathrm{phys}(T = 0)$ through Eq.~(\ref{eq:higgs_stability}). For $T = 0$ we obtain $\alpha_H(T = 0) = (N - 1) \cdot (16 \pi)^{-1} \cdot \lambda_\mathrm{phys}(T = 0)$.

In order to find $\alpha_p(T \gg T_N)$ we need to integrate the thermal RG equations (\ref{eq:TRG_lambda})-(\ref{eq:TRG_mass}) with the initial conditions $\lambda_{\Lambda = \infty} = \lambda_\mathrm{phys}(T = 0)$ and $m^2_{\Lambda = \infty} = m^2_\mathrm{phys}(T = 0)$. Note that the value of $m_\mathrm{phys}^2(T = 0)$ is irrelevant in the limit $T \rightarrow \infty$, where $m_\mathrm{phys}^2(T = 0)/T^2 \rightarrow 0$. The high-temperature parameters are hence determined solely by $\lambda_\mathrm{phys}(T = 0)$ and $T$. In Fig.~\ref{fig:thermal_mass}(a) we plot the ratio of the high- to the low-temperature coupling constant as a function of $\log_{10} \lambda_\mathrm{phys}(T = 0)$. In the weak-coupling regime the effects of the thermal flow are negligible, thus $\lambda_\mathrm{phys}(T = 0)$ and $\lambda_\mathrm{phys}(T \gg T_N)$ can be used interchangeably to formally control the perturbation expansion (although, as we shall show, the flow needs to be included to match quantitatively experimental data). In Fig.~\ref{fig:thermal_mass}(b) the calculated square of the normalized paramagnon mass $m_p^2(T \gg T_N)/T^2$ is shown as a function of the high-temperature coupling constant $\lambda_\mathrm{phys}(T \gg T_N)$ (black solid line).  We observe that $m_p \propto \sqrt{\lambda_\mathrm{phys}(T \gg T_N)} \cdot T$ for $\lambda_\mathrm{phys}(T \gg T_N) \ll 1$. In the weak-coupling limit one obtains analytically $m_p^2 = (N + 2)/12 \cdot \lambda_\mathrm{phys} \cdot T^2$ if the thermal flow of the coupling constant is neglected (red dashed line in Fig.~\ref{fig:thermal_mass}.). The same analytic expression can be derived by a diagrammatic resummation of the infinite class of the ``hard thermal loops'', \cite{PhysRevD45Parwani,PhysRevD53Wang} which indicates the non-perturbative origin of the thermal mass.

Since the masses appear in the denominators of the magnon propagators, the order of some seemingly subleading diagrams is reduced due to the relation $m_p \propto \sqrt{\lambda_\mathrm{phys}(T \gg T_N)}$. The latter leads to reorganization of the perturbation expansion so that the width to mass ratio of the paramagnon at high temperatures, given by Eq.~(\ref{eq:paramagnon_stability}), now takes the form

\begin{align}
\alpha_p = & \frac{\lambda_\mathrm{phys}^2 (N + 2)}{64 \pi} \cdot \frac{T^2}{m_p^2} + O\left(\lambda_\mathrm{phys}^{3/2} \ln \lambda_\mathrm{phys}\right), \label{eq:alpha_p_expanded}
\end{align}

\noindent
where we have made use of the formula $\mathrm{Li}_2(\exp(-x)) = \pi^2/6 + O(x \ln x)$ for $x \rightarrow 0$. The first term on the right-hand-side of Eq.~(\ref{eq:alpha_p_expanded}) is $O(\lambda_\mathrm{phys})$. Since the remainder is of the subleading order $O(\lambda^{3/2}_\mathrm{phys} \ln \lambda_\mathrm{phys})$, it should be discarded (otherwise, one would need to include the higher-loop corrections as well for the sake of consistency). The value of $\alpha_p(T \gg T_N)$ can be now obtained by substituting the calculated paramagnon mass $m_p(T \gg T_N) = m_\mathrm{phys}(T \gg T_N)$ and the coupling constant $\lambda_\mathrm{phys}(T \gg T_N)$ into Eq.~(\ref{eq:alpha_p_expanded}). Note that, since we insert the resummed quantities into the perturbative (two-loop) expression, the latter procedure should be viewed as a variation of renormalized perturbation theory and might be not valid arbitrarily close to the classical transition point. The corrections are, however, expected to be logarithmically small in the renormalized mass scale,\cite{PhysRevLett81Pietroni,PhysLettB35Bergerhoff} hence we are not concerned with them in the present discussion. It now becomes apparent that, due to the emergence of the thermal mass, $\alpha_p(T \gg T_N)$ is of the same order as $\alpha_H(T \rightarrow 0)$, allowing for the linear scaling between these quantities.

\begin{widetext}
\center
\begin{figure}%
\centering
\parbox{0.39\textwidth}{\includegraphics[width = 0.39\textwidth]{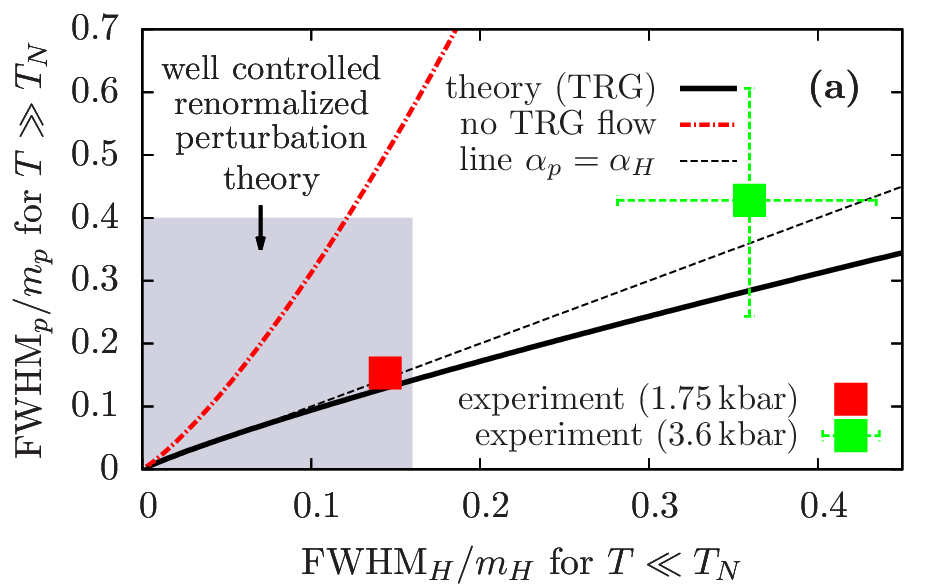}}%
\parbox{0.39\textwidth}{\includegraphics[width = 0.39\textwidth]{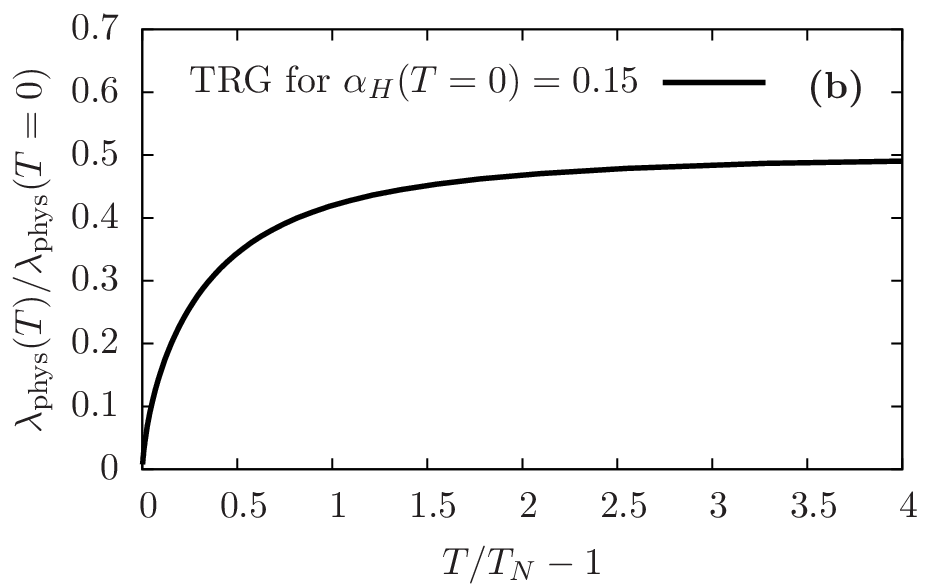}}%

\parbox{0.39\textwidth}{\includegraphics[width = 0.39\textwidth]{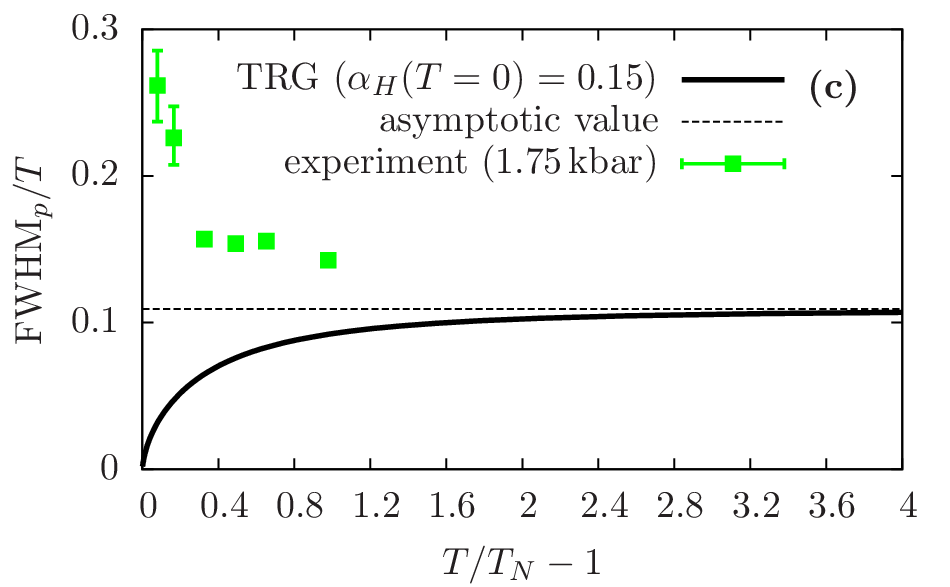}}%
\parbox{0.39\textwidth}{\includegraphics[width = 0.39\textwidth]{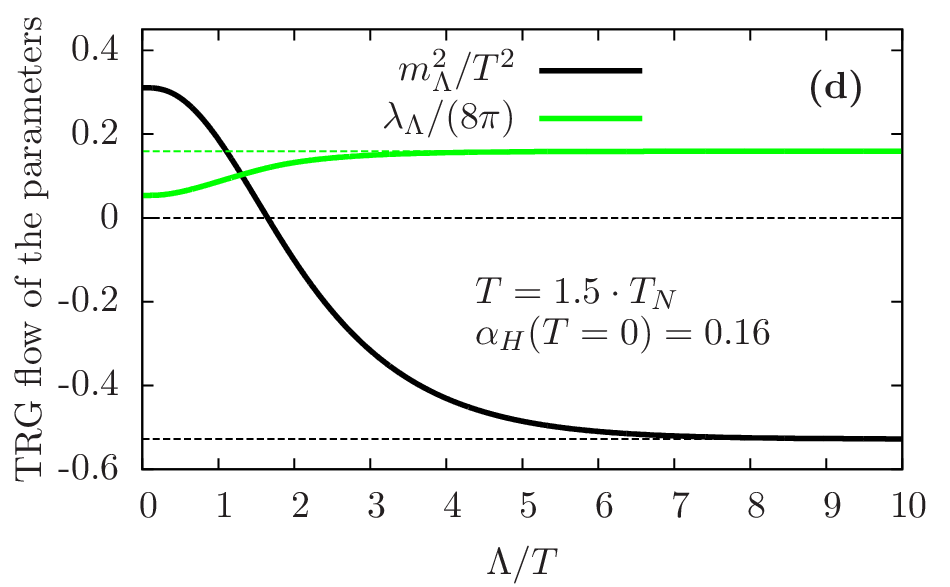}}%
\caption{(Color online) (a) The relation between the stability parameters of the high-temperature paramagnons and the low-temperature amplitude mode. The black solid line is the solution to the thermal renormalization group (TRG) equations and the dashed line $\alpha_p = \alpha_H$ is guide to the eye. Solid squares are experimental data of Ref.~\citenum{NatPhys10Merchant}. The red dot-dashed line shows the relation between $\alpha_p(T \gg T_N)$ and $\alpha_H(T \ll T_N)$ calculated with discarded thermal flow of the coupling constant. The shaded area is defined by the condition $\lambda_\mathrm{phys}(T = 0)/4 < 1$ that marks the regime of applicability of the renormalized perturbation theory. (b) The ratio of the thermal- to the zero-temperature coupling constant as a function of temperature for the parameters chosen so that $\alpha_H(T = 0) = 0.15$. (c) The calculated normalized width $\mathrm{FWHM}_p/T$ of the paramagnons as a function of temperature (solid line) for $\alpha_H(T = 0) = 0.15$ that roughly corresponds to the experimental value for $\mathrm{TlCuCl_3}$ at $p = 1.75\,\mathrm{kbar}$. The green points are the data extracted from Ref.~\citenum{NatPhys10Merchant}. (d) Thermal flow of the parameters for $T = 1.5 \cdot  T_N$ and $\alpha_H(T = 0) = 0.16$. The dashed lines are guides to the eye.} \label{fig:main_result}
\end{figure}
\end{widetext}

The calculated relation between $\alpha_p(T \gg T_N)$ and $\alpha_H(T \ll T_N)$ is depicted in Fig.~\ref{fig:main_result}(a) by a black solid line, which is the main result of the paper. In the regime of stable low-temperature amplitude mode ($\alpha_H \lesssim 0.2$), we get $\alpha_p(T \gg T_N) \approx \alpha_H(T \ll T_N)$ in agreement with experiment (the dashed straight line $\alpha_p = \alpha_H$ is guide to the eye and the solid squares are experimental data for $\mathrm{TlCuCl_3}$). The shaded area in Fig.~\ref{fig:main_result}(a) is defined by the requirement that the effective zero-temperature quartic coefficient $\lambda_\mathrm{phys}(T = 0)/4$ is smaller than one, which ensures applicability of the renormalized perturbation theory based on Eqs.~(\ref{eq:higgs_stability}) and (\ref{eq:alpha_p_expanded}). On general grounds, one expects to fall into this regime sufficiently close to the quantum critical point. Indeed, by inspection of the $T = 0$ RG equation~(\ref{eq:TRG_lambda}) one can see that, for $m_H \rightarrow 0$, $\lambda_\mathrm{phys}(T = 0) \propto 1 / \ln(m_H / \mu_0)$, where $\mu_0$ is a (non-universal) metric factor. In this limit one hence gets $\lambda_\mathrm{phys}(T = 0) \rightarrow 0$ and the computational procedure becomes well controlled. 

In Fig.~\ref{fig:main_result}(a) we also plot the relation between $\alpha_p(T \gg T_N)$ and $\alpha_H(T \ll T_N)$ calculated with neglected thermal flow of the coupling constant, i.e. $\lambda_\mathrm{phys}(T = 0)$ is used in Eq.~(\ref{eq:alpha_p_expanded}) instead of $\lambda_\mathrm{phys}(T \gg T_N)$ (red dot-dashed line). The effect of renormalization of $\lambda_\mathrm{phys}(T)$ is significant in the experimentally accessed parameter range and must be included to match the data. To further illustrate this point, in Fig.~\ref{fig:main_result}(b) we plot the temperature dependence of $\lambda_\mathrm{phys}(T) / \lambda_\mathrm{phys}(T = 0)$ for the the parameters chosen so that $\alpha_H(T = 0) = 0.15$. For $T \gg T_N$ (i.e., in the regime most relevant to the present discussion) the coupling constant saturates at the value reduced relative to $\lambda_\mathrm{phys}(T = 0)$. This downward renormalization is reflected in the value of the paramagnon stability parameter $\alpha_p(T \gg T_N)$ and thereby in the proportionality factor between $\alpha_p(T \gg T_N)$ and $\alpha_H(T \ll T_N)$. As one moves towards the N\'{e}el temperature, the effects of fluctuations become even more pronounced and $\lambda_\mathrm{phys}$ approaches zero for $T \rightarrow T_N$. The latter behavior is an indirect manifestation of critical slowing down, which requires that the paramagnon decay rate $\tau_p^{-1}$ goes to zero as the classical transition is approached.\cite{PhysRevLett81Pietroni,PhysRevD70Zhang} Indeed, from the relation $\tau_p^{-1} \propto \mathrm{FWHM}_p$ and Eq.~(\ref{eq:alpha_p_expanded}) it follows that $\tau_p^{-1} \propto \lambda^2_\mathrm{phys}(T)/m_p(T)\cdot T^2$. Since the paramagnon gap closes ($m_p \rightarrow 0$) for $T \rightarrow T_N$, one arrives at $\lambda_\mathrm{phys} \rightarrow 0$ in this limit.  Critical slowing down can be also seen directly by plotting the normalized paramagnon width $\mathrm{FWHM}_p/T$ vs temperature [black solid line in Fig.~\ref{fig:main_result}(c)]. We observe that the calculated $\mathrm{FWHM}_p/T$ goes to zero for $T \rightarrow T_N$ as anticipated. The solid squares in Fig.~\ref{fig:main_result}(c) are experimental values of $\mathrm{FWHM}_p/T$ for specific pressure $p = 1.75\,\mathrm{kbar}$, extracted from Ref.~\citenum{NatPhys10Merchant}. While the saturation of $\mathrm{FWHM}_p/T$ at high temperatures exhibits the tendency for saturation in agreement with thermal RG prediction, there is a qualitative difference close to the classical as the measured $\mathrm{FWHM}_p/T$ increases close to $T_N$. This point will be addressed in greater detail in the following section.

Finally, a typical thermal RG flow of the parameters is depicted in Fig.~\ref{fig:main_result}(d). The initial conditions at $\Lambda = \infty$ correspond to the N\'{e}el phase ($m_{\Lambda = \infty}^2 < 0$). The thermal effects become significant for $\Lambda \sim 3 \cdot T$ and drive the system to the disordered phase ($m_{\Lambda = 0}^2 = m_\mathrm{phys}^2 > 0$) in the physical limit $\Lambda \rightarrow 0$.

\section{Comparison with experiment: $\boldsymbol{\mathrm{TlCuCl_3}}$ }
\label{section:experiment}

\begin{figure}%
\centering
\parbox{0.39\textwidth}{\includegraphics[width = 0.39\textwidth]{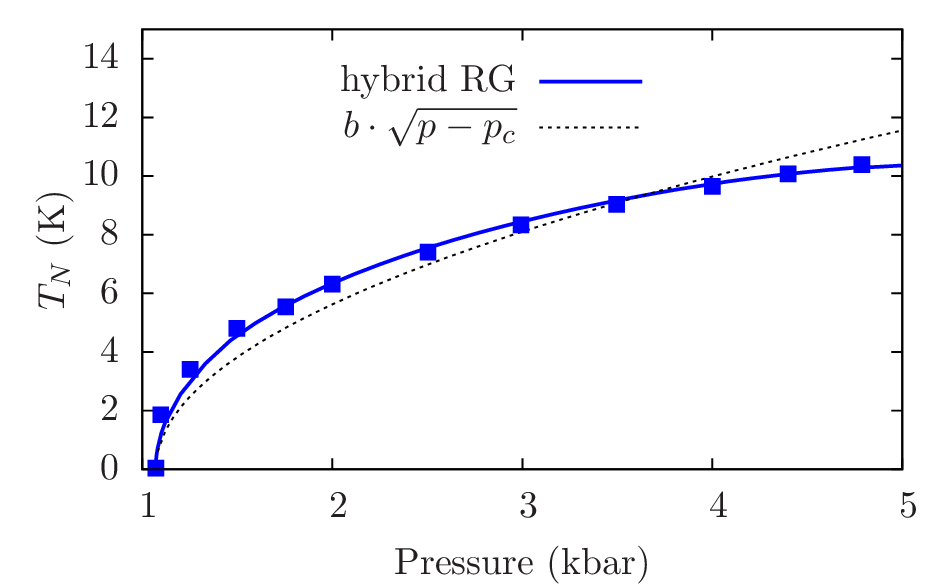}}%
\caption{(Color online) Fit to the experimental data of Ref.~\citenum{NatPhys10Merchant}. (solid squares). Blue line -- hybrid RG. Black dashed line -- fit with a square root form $T_N(p) = b \cdot \sqrt{p - p_c}$.} \label{fig:neel_fit}
\end{figure}

In order to make comparison of the hybrid RG results with the experimental data for the case of a dimerized antiferromagnet $\mathrm{TlCuCl_3}$ across the phase diagram, one needs to address relevant material-specific features, such as easy-plane type magnetic anisotropy $\Delta_\mathrm{an} \approx 0.38 \,\mathrm{meV}$, present in this compound. The latter becomes appreciable if some of the paramagnon masses are smaller than $\Delta_\mathrm{an}$. It happens, e.g., in a narrow slab around the classical transition line, where the anisotropy is expected to induce the crossover from $SU(2)$ to $XY$ behavior. For $\mathrm{TlCuCl_3}$ at $p = 1.75\,\mathrm{kbar}$ the above condition is fulfilled for $T/T_N - 1 \sim 0.4$. Remarkably, below this temperature, the experimental $\mathrm{FWHM}_p/T$ starts to increase [cf. Fig.~\ref{fig:main_result}(c)], which is difficult to reconcile with the anticipated critical slowing down, reproduced by the thermal RG calculation [solid line in Fig.~\ref{fig:main_result}(c)]. Similar behavior has been observed in other anisotropic antiferromagnets, e.g., $S = \frac{5}{2}$ $\mathrm{MnF_2}$ and $\mathrm{Rb_2 Mn F_4}$.\cite{PhysRevB4Schulhof,PhysRevB94Tseng} From the relations $\mathrm{FWHM}_p/T \propto \lambda_\mathrm{phys}^2/m_p \cdot T$ and $\lambda_\mathrm{phys} = \lambda_\Lambda \mid_{\Lambda=0}$ one can see that such an upturn of $\mathrm{FWHM}_p/T$ is consistent with a partial suppression of the thermal RG flow of $\lambda_\Lambda$ below the scale of $\Delta_\mathrm{an}$. Motivated by this observation, instead of systematic inclusion of the anisotropies as new critical variables, we adopt a heuristic approach and cut off the flow of the coupling constant by taking $\mu = \mathrm{max}(|m_\mathrm{phys}^2|^{1/2}, \Delta_\mathrm{an})$ as the renormalization scale and performing a shift  in the infrared-singular polarization loop contributions to the thermal RG equations [$\mathcal{I}_\Lambda''(M^2) \rightarrow \mathcal{I}_\Lambda''(M^2 + \Delta_\mathrm{an}^2)$ in Eqs.~(\ref{eq:TRG_lambda})-(\ref{eq:TRG_mass})]. 

\begin{widetext}

\begin{figure}%
\centering
\parbox{0.33\textwidth}{\includegraphics[width = 0.33\textwidth]{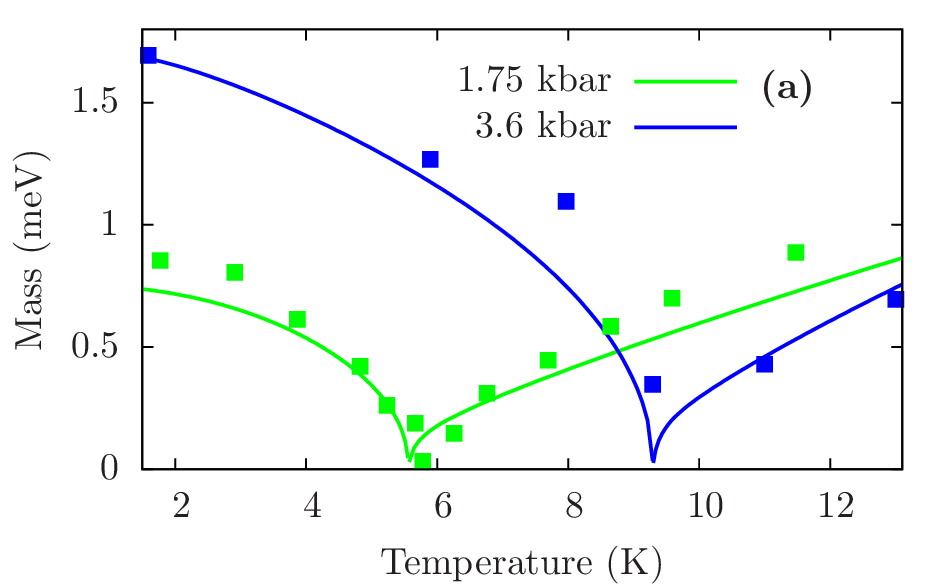}}%
\parbox{0.33\textwidth}{\includegraphics[width = 0.33\textwidth]{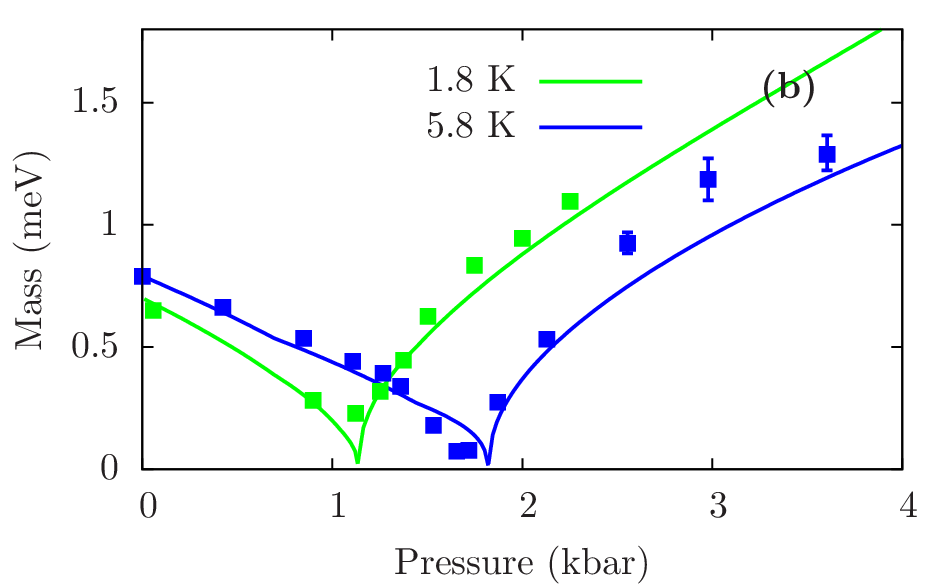}}%
\parbox{0.33\textwidth}{\includegraphics[width = 0.33\textwidth]{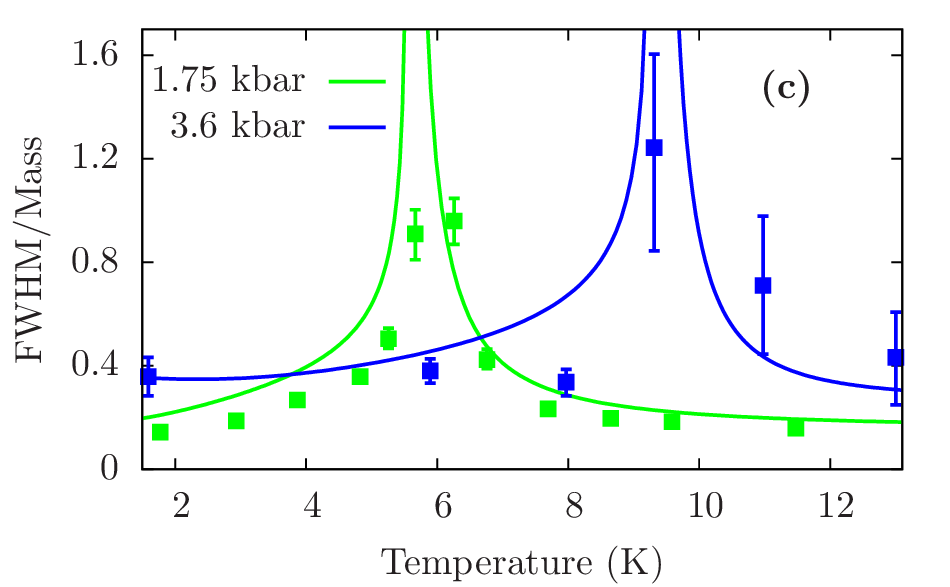}}%

\parbox{0.33\textwidth}{\includegraphics[width = 0.33\textwidth]{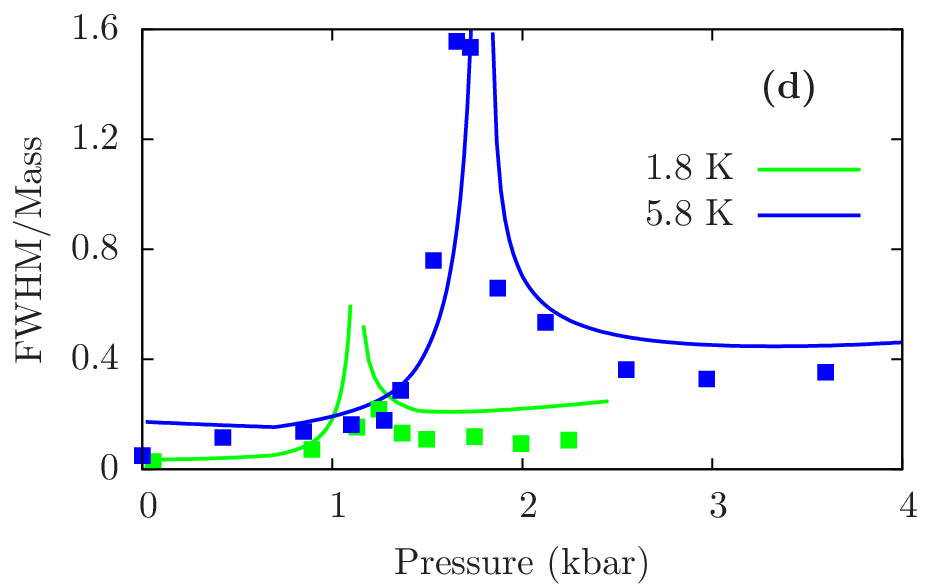}}%
\parbox{0.33\textwidth}{\includegraphics[width = 0.33\textwidth]{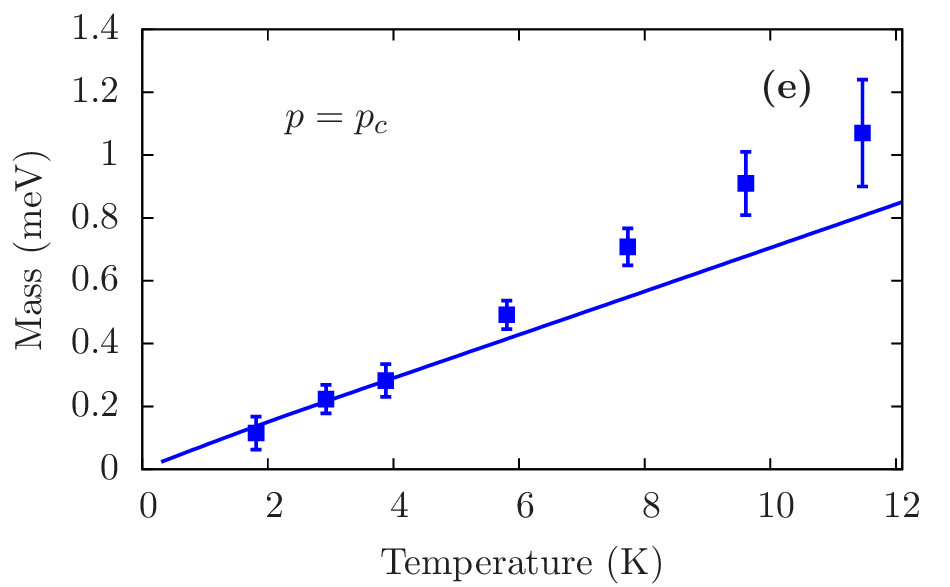}}%
\parbox{0.33\textwidth}{\includegraphics[width = 0.33\textwidth]{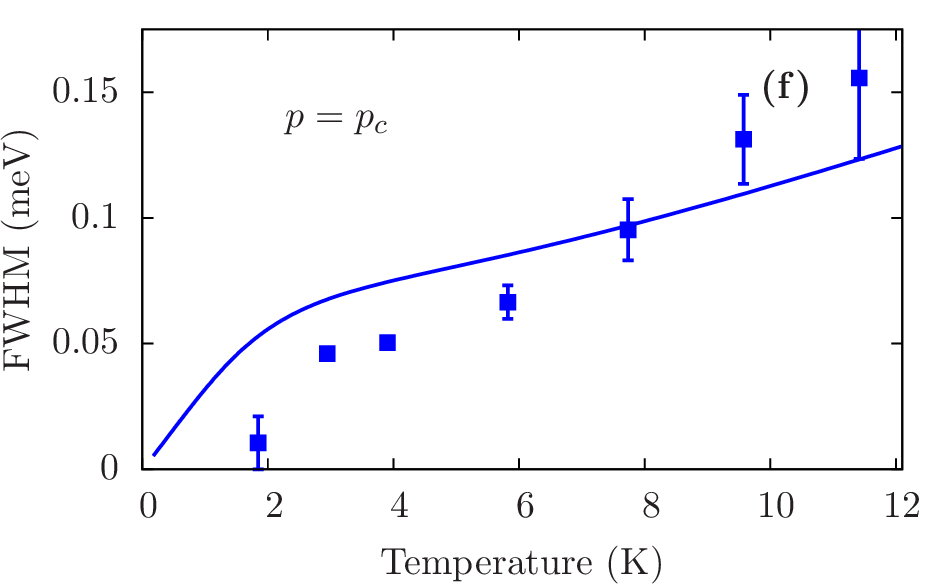}}%
\caption{(Color online) (a)-(d) Temperature- and pressure- dependence of masses and widths of the magnetic excitations in $\mathrm{TlCuCl_3}$. The curves in the low-temperature regime above $p_c = 1.07\,\mathrm{kbar}$ correspond to the amplitude (Higgs) mode, whereas in the high-temperature and low-pressure range to the paramagnons. (e)-(f) Masses and widths of the magnetic excitations at critical pressure as a function of temperature. In all panels the solid lines are solutions to the hybrid RG equations, while the solid squares are the experimental data of Ref.~\citenum{NatPhys10Merchant}. }\label{fig:comparison_with_experiment}
\end{figure}
\end{widetext}

At this point we are fully equipped to make a comparison of the hybrid RG results with experiment and proceed as follows. We define the theory at an arbitrarily chosen scale $\mu_0 = 1\,\mathrm{meV}$ and end up with two free parameters: $m^2_{\mu_0}$ and $\lambda_{\mu_0}$. Close to the critical pressure, we can further expand $m^2_{\mu_0}(p) = a \cdot (p - p_c)$, where $a$ is a numeric coefficient. The two numbers $a$ and $\lambda_{\mu_0}$ are obtained by fitting to the pressure dependence of the N\'{e}el temperature for $\mathrm{TlCuCl_3}$ with the result $a = 0.53\,\mathrm{meV}^2 \mathrm{kbar}^{-1}$ and $\lambda_{\mu_0} = 7.25$. The quality of the fit (blue line in Fig.~\ref{fig:neel_fit}) is remarkable, in contrast to a fit by a simple mean field form $T_N(p) = b \cdot \sqrt{p - p_c}$ that yields $b \approx 5.83\,\mathrm{K} \cdot \mathrm{kbar}^{-1/2}$ (dotted line). Such a non-trivial pressure dependence of the N\'{e}el temperature is likely a manifestation of the logarithmic corrections to scaling at the upper critical dimension, accounted for by the Callan-Symanzik equations.\cite{PhysRevB92Scammell,PhysRevB92Qin} With no other fitting parameters, we can now calculate masses and widths of magnetic excitations across quantum and classical transitions by using the RG equations (\ref{eq:RG_T0_lambda})-(\ref{eq:RG_T0_mass}), (\ref{eq:TRG_lambda})-(\ref{eq:TRG_mass}), combined with Eqs.~(\ref{eq:higgs_stability}) and (\ref{eq:alpha_p_expanded}). The results are confronted with the experimental data of Ref.~\citenum{NatPhys10Merchant} in Fig.~\ref{fig:comparison_with_experiment}.

The temperature dependence of masses of the magnetic excitations for the two selected values of pressure is shown in Fig.~\ref{fig:comparison_with_experiment}(a). At low temperatures the data refer to the amplitude mode in the N\'{e}el phase, while at high temperatures to the paramagnon in the disordered state. The transitions between these phases are marked by closing of the gaps at the pressure-dependent N\'{e}el temperature $T_N(p)$. The same quantities as a function of pressure for fixed temperature, are depicted in Fig.~\ref{fig:comparison_with_experiment}(b).

The corresponding width to mass ratios along $T = \mathrm{const.}$ and $p = \mathrm{const.}$ lines are shown in Figs.~\ref{fig:comparison_with_experiment}(c) and (d), respectively. One can see that the inclusion of the empirical anisotropy scale $\Delta_\mathrm{an}$ has allowed to reconcile the theoretically predicted critical slowing down with the sharp increase of $\alpha_p$ and $\alpha_H$ close to the N\'e{e}l temperature. Moreover, the relevance of residual quantum fluctuations (manifested as corrections to the mean scaling at low-temperatures) can be inferred from Fig.~\ref{fig:comparison_with_experiment}(c). It can be shown by noting that, at the lowest temperatures $T < 2\,\mathrm{K}$, the measured width to mass ratio of the amplitude mode increases by $\sim 100 \%$ as the pressure is varied from $1.75$ to $3.6\,\mathrm{kbar}$ (corresponding to over three-fold change of the distance from the quantum critical point $ p - p_c$). Since in this temperature range $k_B T$ is by an order of magnitude smaller than the amplitude mode energy [cf. Fig.~\ref{fig:comparison_with_experiment}(a)], damping is dominated by the quantum contribution $\alpha_H \approx (N - 1) (16 \pi)^{-1} \cdot \lambda_\mathrm{phys}(T = 0)$. The low-energy behavior of the amplitude mode stability parameter $\alpha_H$ is then controlled by the coupling constant $\lambda_\mathrm{phys}(T = 0)$ which is suppressed at logarithmically slow rate close to the quantum critical point. This relatively weak effect is correctly reproduced by the hybrid RG that also quantitatively agrees with experiment [cf. Fig.~\ref{fig:comparison_with_experiment}(c)]. Solving both $T = 0$ RG equations (\ref{eq:RG_T0_lambda})-(\ref{eq:RG_T0_mass}) and thermal RG equations (\ref{eq:TRG_lambda})-(\ref{eq:TRG_mass}) is hence necessary to match  the data across the phase diagram.

Finally, in Figs.~\ref{fig:comparison_with_experiment}(e)-(f) the temperature dependence of the full widths and masses of the magnetic excitations is presented as a function of temperature at the critical pressure $p_c \approx 1.07\,\mathrm{kbar}$. The hybrid RG solution for the paramagnon mass obeys a linear-$T$ scaling and agrees quantitatively with experiment at low temperatures. The width of the paramagnon, however, deviates from linear $T$-dependence and exhibits a hump close to $T = 4\,\mathrm{K}$. We attribute this behavior to the spin anisotropy $\Delta_\mathrm{an}$ which remains non-zero across the phase transition and becomes comparable with the thermal mass around this temperature. Interestingly, a weak feature can be seen seen in the experimental data at $\sim 3\,\mathrm{K}$, but detailed analysis cannot be performed due to small numbers of available data points. Worse agreement of the thermal RG result with experiment in Fig.~\ref{fig:comparison_with_experiment}(f) relative to Figs.~\ref{fig:comparison_with_experiment}(a)-(e) is likely a consequence of approximate inclusion of magnetic anisotropy in our analysis.

\section{Summary and discussion}
\label{sec:summary}

We have proposed a mechanism leading to a linear scaling (with proportionality factor $\approx 1$) between the width to mass ratios of the high-temperature paramagnon and the low-temperature amplitude (Higgs) mode in a dimerized antiferromagnet $\mathrm{TlCuCl_3}$, based on: (i) reorganization of the perturbation expansion by the emergence of the thermal mass $\sim \sqrt{\lambda_\mathrm{phys}} \cdot T$ and (ii) sizable thermal flow of the effective mutli-magnon coupling $\lambda_\mathrm{phys}$.

The hybrid Callan-Symanzik + thermal RG approach has allowed us to include both characteristics of the quantum critical point (such as logarithmic suppression of the zero-temperature coupling constant) and the classical critical point (e.g., critical slowing down close to the classical transition). These aspects of the hybrid RG approach, after inclusion of the empirical anisotropy scale, have made it possible to perform a successful comparison with experimental data for $\mathrm{TlCuCl_3}$ across the phase diagram in the pressure-temperature plane.

At the end, we would like to point out that there exist alternative approaches to damped excitations, implicitly involving resummation of subclasses of higher-order processes. One of them is based on substitution of ``dressed'', rather than bare, propagators into the expressions for the dynamical structure factors of magnetic modes. This approach has been employed in the context of the longitudinal spin fluctuations in iron pnictides,\cite{EurPhysJB89Fidrysiak} and extended to a dimerized antiferromagnet\cite{arxivScammell2016} which yields a good agreement with the experimental data for $\mathrm{TlCuCl_3}$ in the disordered phases for $p > p_c$ and $p = 0\,\mathrm{kbar}$, but fails in the intermediate regime $0 < p < p_c$. The hybrid renormalization group, used here, agrees with the data reasonably well in the disordered state for both $p > p_c$ and $0 < p < p_c$, but overestimates $\mathrm{FWHM}_p$ for the lowest pressure $p = 0\,\mathrm{kbar}$ at $T = 5.8\,\mathrm{K}$ (cf. Figs.~\ref{fig:comparison_with_experiment}(a)-(d) and the subsequent discussion in Sec.~\ref{section:experiment}).

\begin{acknowledgments}
The work was supported by Grant MAESTRO No.~DEC-2012/04/A/ST3/00342 from the National Science Centre (NCN).
\end{acknowledgments}

\appendix

\section{Widths of the magnetic excitations} \label{appendix:widths_of_excitations}

In this Appendix all the calculations are performed using resummed parameters $m^2_\mathrm{phys}$ and $\lambda_\mathrm{phys}$, but we omit the subscripts for brevity.

\subsection{Amplitude mode below $\boldsymbol{T_N}$} 

The leading order process, giving rise to the amplitude mode decay, is given by the diagram of Fig.~\ref{fig:decay_processes}(b) whose contribution to the mass operator $\Sigma_{H}(i \omega_n, \mathbf{k})$ reads

\begin{align}
\Sigma_{H} = &\lambda m_H^2 (N - 1) \cdot T \SumInt \limits_m \frac{d^3 \mathbf{q}}{(2 \pi)^3} G_{\mathrm{TM}}(i \omega_{n - m}, \mathbf{k} - \mathbf{q}) \times \nonumber \\ & \times G_{\mathrm{TM}}(i \omega_m, \mathbf{q}), 
\end{align}

\noindent
where $\omega_n = 2 \pi n / T$ are bosonic Matsubara frequencies and $G_{\mathrm{TM}}(i \omega_n, \mathbf{k}) = (\omega_n^2 + \mathbf{k}^2)^{-1}$ denotes the propagator of the spin-wave mode.

The imaginary part of the real-time mass operator at the magnetic zone center $\Sigma^{''}_{H}(E, \mathbf{0})$  can be evaluated by switching to the real-time representation 

\begin{align}
\frac{1}{\omega_n^2 + E_\mathbf{k}^2} = \int \limits_0^{1/T} d\tau \frac{\mathrm{e}^{i \omega_n \tau} }{2 E_\mathbf{k}} \cdot \left[ \mathrm{e}^{\tau E_\mathbf{k}} n_\mathbf{k} + \mathrm{e}^{-\tau E_\mathbf{k}} (1 + n_\mathbf{k}) \right],
\end{align}

\noindent
and performing analytic continuation $i \omega_n \rightarrow E + i\epsilon$ after working out the integrals over $\tau$ variables. The full width of the amplitude mode $\mathrm{FWHM}_H$ is then evaluated as

\begin{align}
\mathrm{FWHM}_H \equiv & \Sigma^{''}_{H}(m_H, \mathbf{0}) / m_H = \pi \lambda m_H (N - 1) \times \nonumber \\ & \times \int \frac{d^3 \mathbf{q}}{(2\pi)^3} \delta(m_H - 2 E_\mathbf{q}) \cdot \frac{1 + 2 n_\mathbf{q}}{(2 E_\mathbf{q})^2} = \nonumber \\ & = \frac{\lambda m_H (N - 1)}{16 \pi} \cdot [1 + 2\cdot  n(m_H/2)],
\end{align}

\noindent
from which Eq.~(\ref{eq:higgs_stability}) follows.

\subsection{Paramagnons above $\boldsymbol{T_N}$}

The contribution to the paramagnon mass operator from the diagram, shown in Fig.~\ref{fig:decay_processes}(a), is given by

\begin{align}
\Sigma_{p} = &2 \lambda^2 (N + 2) \cdot T \SumInt \limits_m \frac{d^3 \mathbf{q}_1}{(2 \pi)^3} \frac{d^3 \mathbf{q}_2}{(2 \pi)^3} G_{p}(i \omega_{n - m}, \mathbf{k} - \mathbf{q}_1) \times \nonumber \\ & \times G_{p}(i \omega_{m-l}, \mathbf{q}_1 - \mathbf{q}_2) \cdot G_{p}(i \omega_{l}, \mathbf{q}_2), 
\end{align}

\noindent
where $G_{p}(i \omega_n, \mathbf{k}) = (\omega_n^2 + \mathbf{k}^2 + m_p^2)^{-1}$. By switching to the real-time representation and performing analytic continuation $i \omega_n \rightarrow E + i\epsilon$, we arrive at $\Sigma_p''(E, \mathbf{0}) \equiv \Sigma_p^{(1) ''}(E, \mathbf{0}) + \Sigma_p^{(2) ''}(E, \mathbf{0})$, where

\begin{align}
\Sigma^{(1) ''}_{p} =&  3 \int \frac{d^3 \mathbf{q}_1}{(2 \pi)^3} \frac{d^3 \mathbf{q}_2}{(2 \pi)^3} \frac{\delta(E - E_{\mathbf{q}_1} - E_{\mathbf{q}_1} + E_{\mathbf{q}_1 - \mathbf{q}_2})}{(2 E_{\mathbf{q}_1}) \cdot (2 E_{\mathbf{q}_2}) \cdot (2 E_{\mathbf{q}_1 - \mathbf{q}_2}) } \times \nonumber \newline \\ &\times \left[ \mathrm{e}^{(- E_{\mathbf{q}_1} - E_{\mathbf{q}_1} + E_{\mathbf{q}_1 - \mathbf{q}_2})/T} - 1\right] \cdot (1 + n_{\mathbf{q}_1}) \times \nonumber \newline \\ &\times (1 + n_{\mathbf{q}_2})  \cdot n_{\mathbf{q}_1 - \mathbf{q}_2} \cdot 2 \lambda^2 (N + 2), \label{eq_app:sigma_prim_1}\\
\Sigma^{(2) ''}_{p} =&  \int \frac{d^3 \mathbf{q}_1}{(2 \pi)^3} \frac{d^3 \mathbf{q}_2}{(2 \pi)^3} \frac{\delta(E - E_{\mathbf{q}_1} - E_{\mathbf{q}_1} - E_{\mathbf{q}_1 - \mathbf{q}_2})}{(2 E_{\mathbf{q}_1}) \cdot (2 E_{\mathbf{q}_2}) \cdot (2 E_{\mathbf{q}_1 - \mathbf{q}_2}) } \times \nonumber \newline \\ &\times \left[ \mathrm{e}^{(- E_{\mathbf{q}_1} - E_{\mathbf{q}_1} - E_{\mathbf{q}_1 - \mathbf{q}_2})/T} - 1\right] \cdot (1 + n_{\mathbf{q}_1}) \times \nonumber \newline \\ &\times (1 + n_{\mathbf{q}_2}) \cdot (1 + n_{\mathbf{q}_1 - \mathbf{q}_2}) \cdot 2 \lambda^2 (N + 2). \label{eq_app:sigma_prim_2}
\end{align}

\noindent
for $E > 0$. By inspecting the arguments of the Dirac delta functions in Eqs.~(\ref{eq_app:sigma_prim_1})-(\ref{eq_app:sigma_prim_2}) one can see that the first term $\Sigma_p^{(1) ''}$ involves paramagnons from the thermal bath and hence is non-zero for $E = m_p$, while $\Sigma_p^{(2) ''}$ is the three-paramagnon decay process process which vanishes for $E < 3 m_p$. It is then sufficient to calculate $\Sigma_p^{(1) ''}$. 

By introducing dimensionless variables $x = E_{\mathbf{q}_1}/T$ and $y = E_{\mathbf{q}_2}/T$ we arrive at the formula

\begin{align}
\Sigma^{(1) ''}_{p}&(m_p, \mathbf{0}) =  \frac{3 \lambda^2 (N + 2)}{32 \pi^3} \cdot T^2 \cdot (1 - \mathrm{e}^{-m_p/T}) \times \nonumber \\ &\times \int \limits_{\frac{m_p}{T}}^\infty dx \int \limits_{\frac{m_p}{T}}^\infty dy \frac{1}{\mathrm{e}^{-x} - 1} \frac{1}{\mathrm{e}^{-y} - 1} \frac{1}{\mathrm{e}^{x + y - m_p/T} - 1} = \nonumber \\ & = \frac{3\lambda^2 (N + 2)}{32 \pi^3} \cdot T^2 \cdot \mathrm{Li}_2(\mathrm{e}^{-m_p/T}),
\end{align}

\noindent
where $\mathrm{Li}_2(x)$ is the dilogarithm. The full width of the paramagnon then reads $\mathrm{FWHM}_p = \Sigma^{(1)''}_{p}(m_p, \mathbf{0})/m_p$ and Eq.~(\ref{eq:paramagnon_stability}) is thus reproduced.

\section{Derivation of the RG equations}
\label{appendix:RG_equations}

We start with the Lagrangian, given by Eq.~(\ref{eq:lagrangian}). The model can be discussed in a unified manner both at $T = 0$ and $T > 0$ in terms of the quantum effective action, which is a generator of the one-particle irreducible vertex functions.\cite{BookZuber} At the one-loop level, the latter may be computed by performing the shift $\boldsymbol{\varphi} \rightarrow \boldsymbol{\varphi}_\mathrm{cl} + \delta\boldsymbol{\varphi}$, where $\boldsymbol{\varphi}_\mathrm{cl}$ is the classical field, and integrating out the fluctuations quadratic in $\delta \boldsymbol{\varphi}$. One obtains

\begin{align}
\mathcal{S}_\mathrm{eff} =  \mathcal{S}[\boldsymbol{\varphi}_\mathrm{cl}] + & \frac{1}{2} \mathrm{Tr} \ln \{ (-\Delta + m^2) \delta^{\alpha \beta} + \nonumber \\ + & \lambda \boldsymbol{\varphi}^2_\mathrm{cl} [P_\perp^{\alpha \beta} + 3 P_{||}^{\alpha \beta}] \}, 
\end{align}

\noindent
where $\mathcal{S}(\boldsymbol{\varphi}_\mathrm{cl}) = \int_0^{1/T} d\tau \int d^3 x \mathcal{L}(\boldsymbol{\varphi}_\mathrm{cl})$, and $P_{||}^{\alpha \beta} = \varphi_\mathrm{cl}^\alpha \varphi_\mathrm{cl}^\beta / \boldsymbol{\varphi}_\mathrm{cl}^2$, $P_{\perp}^{\alpha \beta} = \delta^{\alpha \beta} - P_{||}^{\alpha \beta}$ are projectors onto the directions parallel and perpendicular to the classical field $\boldsymbol{\varphi}_\mathrm{cl}$, respectively. The effective potential for a constant field ${\varphi}^i_\mathrm{cl} = F \delta^{i, N}$ is then given by $\mathcal{V}_\mathrm{eff}(F) = \frac{T}{V} \cdot \mathcal{S}_\mathrm{eff}(\boldsymbol{\varphi}_\mathrm{cl})$, where $V$ is the volume of space.

We identify the physical temperature-dependent mass parameter $m_\mathrm{phys}(T)$ and coupling constant $\lambda_\mathrm{phys}(T)$ with the proper derivatives of the effective potential, taken at its minimum, i.e.,

\begin{align}
m^2_\mathrm{phys} = & \frac{\partial^2 \mathcal{V}_\mathrm{eff}}{\partial F^2} - \frac{1}{2} F^2 \frac{\partial^4 \mathcal{V}_\mathrm{eff}}{\partial F^4} =  m^2 + \delta m^2 \nonumber + \\& + \lambda [(N - 1) \mathcal{A}(m^2 + \lambda F^2) \nonumber + 3 \mathcal{A}(m^2 + 3 \lambda F^2)] \nonumber  \\ & - \lambda^2 F^2 [(N - 1) \mathcal{B}(m^2 + \lambda F^2) \nonumber + \\ & + 9 \mathcal{B}(m^2 + 3 \lambda F^2)], \label{eq_app:m2_definition} \\
\lambda_\mathrm{phys} =  &\frac{1}{6} \frac{\partial^4 \mathcal{V}_\mathrm{eff}}{\partial F^4} = \lambda + \delta \lambda + \lambda^2 [(N - 1) \mathcal{B}(m^2 + \lambda F^2) + \nonumber \nonumber \\ & + 9 \mathcal{B}(m^2 + 3 \lambda F^2)], \label{eq_app:lambda_definition} 
\end{align}

\noindent
where

\begin{align}
\mathcal{A}(M^2) =& \mathcal{A}_0(M^2) + \int  \limits_0^\infty \frac{{d}k k^2}{2 \pi^2} \frac{n(\sqrt{M^2 + k^2})}{\sqrt{M^2 + k^2}}, \label{eq_app:A_definition}\\
\mathcal{B}(M^2) = &\mathcal{B}_0(M^2) + \int \limits_0^\infty \frac{dk k^2}{2 \pi^2} \frac{{d}}{{d} M^2} \frac{n(\sqrt{M^2 + k^2})}{\sqrt{M^2 + k^2}},\label{eq_app:B_definition}\\
\mathcal{A}_0(M^2) =& \mu^{2 \epsilon} \int \frac{d^{4 - 2\epsilon} \mathbf{k}}{(2 \pi)^{4 - 2\epsilon}} \frac{1}{\mathbf{k}^2 + M^2} = - \frac{M^2}{(4 \pi)^2} \times \nonumber \\ \times& \left(\frac{1}{\epsilon} - \gamma + \ln(\frac{\mu^2}{M^2}) + \ln(4 \pi) + 1 + O(\epsilon)\right),  \label{eq_app:A0_definition} \\
\mathcal{B}_0(M^2) =& -\mu^{2 \epsilon} \int \frac{d^{4 - 2\epsilon} \mathbf{k}}{(2 \pi)^{4 - 2\epsilon}} \frac{1}{(\mathbf{k}^2 + M^2)^2} = - \frac{1}{(4 \pi)^2} \times \nonumber \\ \times& \left(\frac{1}{\epsilon} - \gamma + \ln(\frac{\mu^2}{M^2}) + \ln(4 \pi) + O(\epsilon)\right), \label{eq_app:B0_definition}
\end{align}

\noindent
and $\gamma \approx 0.5772$ denotes the Euler's constant. The term $-1/2 \cdot F^2 \cdot \partial^2 \mathcal{V}_\mathrm{eff} / \partial F^2$ in the definition of $m^2_\mathrm{phys}$ [Eq.~(\ref{eq_app:m2_definition})] is to cancel the trivial shift coming from the quartic interaction $\lambda \boldsymbol{\varphi}^4$ in the ordered phase ($F^2 > 0$). The integrals $\mathcal{A}(M^2)$ and $\mathcal{B}(M^2)$ [Eqs.~(\ref{eq_app:A_definition})-(\ref{eq_app:B_definition})] have been split into the zero temperature parts $\mathcal{A}_0$ and $\mathcal{B}_0$, and finite-temperature remainders. The short-range divergences are contained only in $\mathcal{A}_0$ and $\mathcal{B}_0$ which have been computed by dimensional regularization with $\epsilon = 2 - D/2$ and $\mu$ being the regularization scale.

To this point, the expressions (\ref{eq_app:m2_definition})-(\ref{eq_app:lambda_definition}) are equivalent to the environment-dependent resummation scheme, developed in Ref.~\citenum{PhysRevD71Jakovac}. In the following subsections we describe renormalization group improvement of this procedure.

\subsection{Callan-Symanzik equations}

Since at finite temperatures, no new divergences are generated, one can first set $T = 0$ (or, equivalently, introduce the cutoff to the Bose factors $n(E_\mathbf{k}) \rightarrow n(E_\mathbf{k}) \times \theta(|\mathbf{k}| - \Lambda)$ and take the limit $\Lambda \rightarrow \infty$). The $1/\epsilon$ terms, divergent for $D \rightarrow 4$, are now cancelled by the counter-terms

\begin{align}
\frac{\delta m^2}{m^2} =& \frac{\lambda (N + 2)}{(4 \pi)^2} \cdot  \left(\frac{1}{\epsilon} - \gamma + \ln(4\pi) + 1\right), \\ \delta \lambda =& \frac{\lambda^2 (N + 2)}{(4 \pi)^2} \cdot \left(\frac{1}{\epsilon} - \gamma + \ln(4\pi)\right)
\end{align}

\noindent
so that the limit $\epsilon \rightarrow 0$ can be taken. Note that the form of countertems is the same for $F = 0$ (disordered phase) and $F^2 >0$ (ordered phase). 

For $F = 0$ and $T = 0$ one arrives then at the simple expressions

\begin{align}
m^2_\mathrm{phys} = &\frac{d^2 \mathcal{V}_\mathrm{eff}}{d F^2} = m^2 - \frac{2 (N + 2) \lambda}{(4 \pi)^2} \cdot m^2 \cdot \ln(\frac{\mu}{m}), \label{eq_app:m2_disordered_expanded} \\ 
\lambda_\mathrm{phys} = & \frac{1}{6} \frac{\partial^4 \mathcal{V}_\mathrm{eff}}{\partial F^4} =  \lambda - \frac{2 (N + 8) \lambda^2}{(4 \pi)^2} \cdot \ln(\frac{\mu}{m}). \label{eq_app:lambda_disordered_expanded}
\end{align}

Equations~(\ref{eq:RG_T0_lambda})-(\ref{eq:RG_T0_mass}) can now be obtained from the requirement that the bare vertex functions do not depend on the scale $\mu$. Technically it can be achieved by requiring that the total derivative of the right-hand-side of Eqs.~(\ref{eq_app:m2_disordered_expanded})-(\ref{eq_app:lambda_disordered_expanded}) is zero to the leading order in $\lambda$. Moreover, one can also see that $m^2_\mathrm{phys} = m^2_\mu \mid_{\mu = m_\mathrm{phys}}$ and $\lambda_\mathrm{phys} = \lambda_\mu \mid_{\mu = m_\mathrm{phys}}$.

In the ordered phase the discussion is more subtle as Eqs.~(\ref{eq_app:m2_definition})-(\ref{eq_app:lambda_definition}) become spoiled by infrared divergences arising from the massless Goldstone modes. This problem can be overcome by noting that the physical parameters, controlling the amplitude mode, are defined by vertex functions at energy scale $E \sim m_H \sim \sqrt{2 |m_\mathrm{phys}^2|} $ rather than at $E=0$ as implicitly encoded in the effective potential (this issue is not essential in the disordered phase, where the divergences are suppressed by the paramagnon mass). Taking that into account  would effectively cut off the contribution from the Goldstone modes at $E \sim m_H$. Here we adopt the formulas $m^2_\mathrm{phys}(T = 0) = m^2_\mu \mid_{\mu = \sqrt{|m_\mathrm{phys}^2|}}$ and $\lambda_\mathrm{phys}(T = 0) = \lambda_\mu \mid_{\mu = \sqrt{|m_\mathrm{phys}^2|}}$ for both ordered and disordered phase.

\subsection{Thermal renormalization group}

We now recover the finite-temperature physics by going from $\Lambda = \infty$ to $\Lambda = 0$ in the Wilson renormalization group sense. By integrating out thermal fluctuations with wavevectors $|\mathbf{k}|$ in the infinitesimal slabs $(\Lambda, \Lambda + d\Lambda)$ and taking into account the flow of the parameters through the process, from Eqs.~(\ref{eq_app:m2_definition})-(\ref{eq_app:B0_definition}), we get

\begin{align}
\lambda_{\Lambda + d\Lambda} =& \lambda_\Lambda + (N - 1) \cdot \lambda^2_\Lambda \cdot \mathcal{I}_\Lambda^{''}(m_{\perp, \Lambda}^2) \cdot \frac{d\Lambda}{\Lambda} \nonumber \\ &+ 9 \lambda^2_\Lambda \cdot \mathcal{I}_\Lambda^{''}(m_{||, \Lambda}^2) \cdot \frac{d\Lambda}{\Lambda}, \label{eq_app:pre_TRG_lambda}  
\end{align}

\noindent
and

\begin{align}
 m^2_{\Lambda + d\Lambda} = & m^2_\Lambda +(N - 1) \lambda_\Lambda \cdot \mathcal{I}_\Lambda^{'}(m_{\perp, \Lambda}^2) \cdot \frac{d\Lambda}{\Lambda} \nonumber \\ &+  3 \lambda_\Lambda \cdot \mathcal{I}_\Lambda^{'}(m_{||, \Lambda}^2) \cdot \frac{d\Lambda}{\Lambda} - 9 \lambda^2_\Lambda F^2_\Lambda \cdot \mathcal{I}_\Lambda^{''}(m_{||, \Lambda}^2) \cdot \frac{d\Lambda}{\Lambda} \nonumber \\ &- (N - 1) \lambda^2_\Lambda F^2_\Lambda \cdot \mathcal{I}_\Lambda^{''}(m_{\perp, \Lambda}^2) \cdot \frac{d\Lambda}{\Lambda},  \label{eq_app:pre_TRG_mass}
\end{align}

\noindent
where $\mathcal{I}_\Lambda'$ and $\mathcal{I}_\Lambda''$ are given by Eqs.~(\ref{eq:I'})-(\ref{eq:I''}). Equations~(\ref{eq_app:pre_TRG_lambda})-(\ref{eq_app:pre_TRG_mass}) can be now transformed into Eqs.~(\ref{eq:TRG_lambda})-(\ref{eq:TRG_mass}).

\bibliography{bibliography}

\end{document}